\documentclass{article}

\usepackage{microtype}
\usepackage{graphicx}
\usepackage{subcaption}
\usepackage{booktabs} 

\usepackage{hyperref}

\usepackage[accepted]{icml2025}

\usepackage{amsmath}
\usepackage{amssymb}
\usepackage{mathtools}
\usepackage{amsthm}

\usepackage[capitalize,noabbrev]{cleveref}

\usepackage[utf8]{inputenc} 
\usepackage[T1]{fontenc}    
\usepackage{url}            
\usepackage{amsfonts}       
\usepackage{nicefrac}       
\usepackage{xcolor}         
\graphicspath{{figures/}}
\usepackage{multirow}

\theoremstyle{plain}

\theoremstyle{definition}

\theoremstyle{remark}

\usepackage[textsize=tiny]{todonotes}

\icmltitlerunning{Diversity by Design: Addressing Mode Collapse Improves scRNA-seq Perturbation Modeling on Well-Calibrated Metrics}

\begin{document}

\twocolumn[
\icmltitle{Diversity by Design: Addressing Mode Collapse Improves scRNA-seq Perturbation Modeling on Well-Calibrated Metrics}



\icmlsetsymbol{equal}{*}

\begin{icmlauthorlist}
\icmlauthor{Gabriel M. Mejia}{equal,comp}
\icmlauthor{Henry E. Miller}{equal,comp}
\icmlauthor{Francis J. A. Leblanc}{comp}
\icmlauthor{Bo Wang}{sch}
\icmlauthor{Brendan Swain}{comp}
\icmlauthor{Lucas Paulo de Lima Camillo}{comp}
\end{icmlauthorlist}

\icmlaffiliation{comp}{Shift Bioscience, Cambridge, UK}
\icmlaffiliation{sch}{University of Toronto, Vector Institute, Toronto, Canada}

\icmlcorrespondingauthor{Lucas Paulo de Lima Camillo}{lucas@shiftbioscience.com}

\icmlkeywords{Virtual Cell, Foundation Model, Genetic Perturbation}

\vskip 0.3in
]



\printAffiliationsAndNotice{\icmlEqualContribution} 

\begin{abstract}
Recent benchmarks reveal that models for single-cell perturbation response are often outperformed by simply predicting the dataset mean. We trace this anomaly to a metric artifact: control-referenced deltas and unweighted error metrics reward mode collapse whenever the control is biased or the biological signal is sparse. Large-scale \textit{in silico} simulations and analysis of two real-world perturbation datasets confirm that shared reference shifts, not genuine biological change, drives high performance in these evaluations. We introduce differentially expressed gene (DEG)–aware metrics, weighted mean-squared error (WMSE) and weighted delta $R^{2}$ ($R^{2}_{w}(\Delta)$) with respect to all perturbations, that measure error in niche signals with high sensitivity. We further introduce negative and positive performance baselines to calibrate these metrics. With these improvements, the mean baseline sinks to null performance while genuine predictors are correctly rewarded. Finally, we show that using WMSE as a loss function reduces mode collapse and improves model performance.
\end{abstract}



\section{Introduction}
\label{introduction}

\begin{figure}[ht!]
\centering
\includegraphics[width=0.99\linewidth]{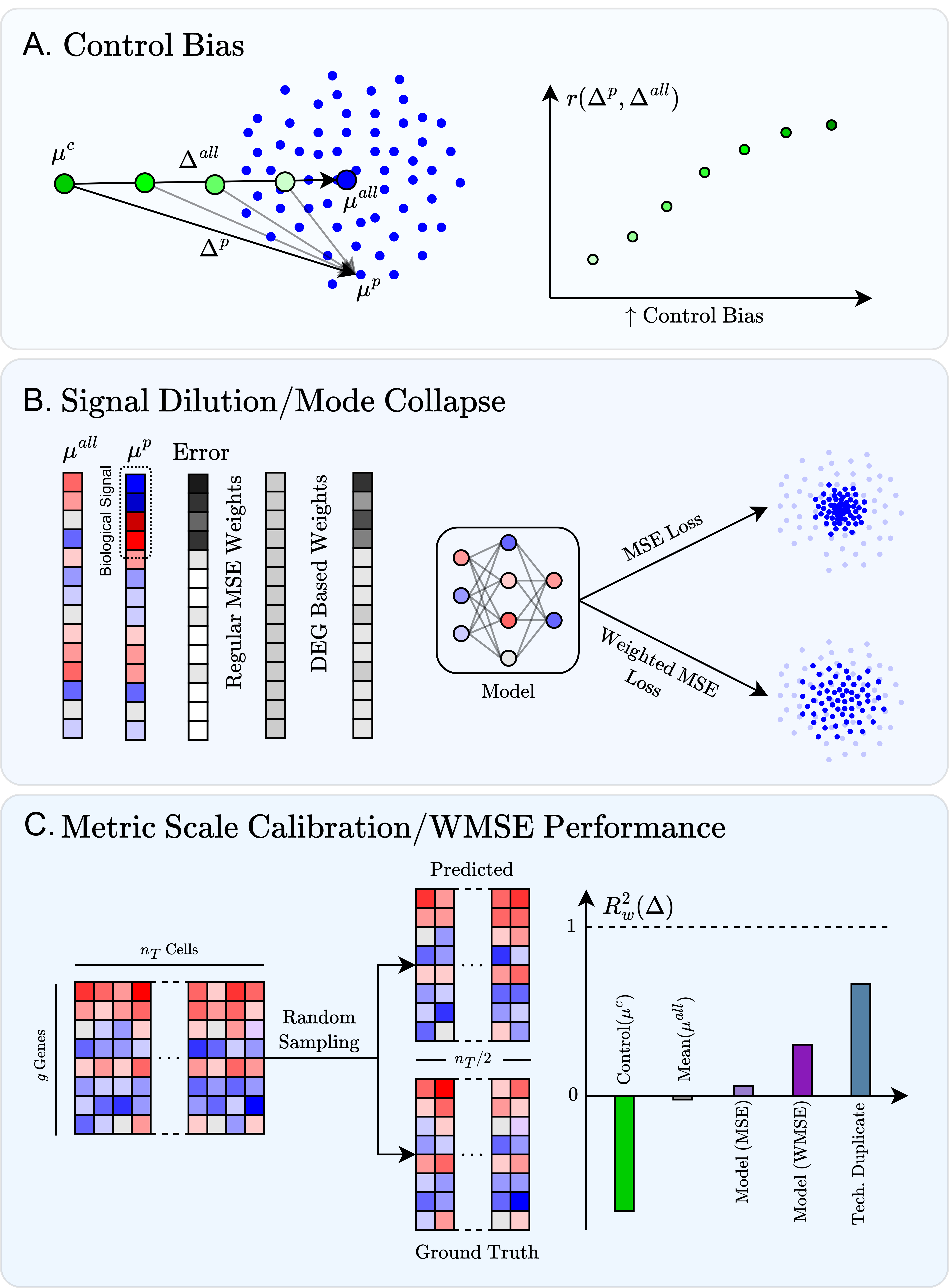}
\caption{(A) Under increasing amounts of systematic control bias, the Pearson correlation between $\Delta^p$ and $\Delta^{all}$ increases artificially. (B) True biological signal is diluted in scRNA-seq data causing mode collapse in model predictions. Introducing biologically aware weights (WMSE) ameliorates this problem. (C) We introduce metrics with a performance scale calibrated by negative and positive baselines. We also find that WMSE as a training objective improves model performance compared to MSE.}
\label{fig:Graphical-abstract}
\phantomsubcaption\label{fig:Graphical-abstract:a}
\phantomsubcaption\label{fig:Graphical-abstract:b}
\phantomsubcaption\label{fig:Graphical-abstract:c}
\vskip -0.3in
\end{figure}

In recent years, technological breakthroughs in experimental methodologies have catalyzed the emergence of large-scale, publicly available, single-cell RNA sequencing (scRNA-seq) perturbation datasets \cite{peidli2024scperturb}, which capture phenotypic changes of individual cells under specific perturbations. Models trained on these datasets to predict perturbation responses may unlock virtual molecule and genetic perturbation screening capabilities, which could yield novel therapeutics that reverse disease and restore cell function. In recent years, researchers have proposed a diverse array of perturbation-response models leveraging various learning paradigms, including optimal transport (CellOracle \cite{kamimoto2023dissecting}), prior knowledge graph learning (GEARS \cite{roohani2024predicting}), and transformer-based foundation models (scGPT \cite{cui2024scgpt}, scFoundation \cite{hao2024large}).

However, recent benchmarking studies have reported that naïve predictors such as linear models often outperform these sophisticated architectures \cite{li2024benchmarking, wu2024perturbench, csendes2025benchmarking, li2024systematic, bendidi2024benchmarking, wenteler2024perteval, ahlmann2024deep}. Even more concerning, a mean baseline, consisting of predicting the average of all perturbed cells in the training set (disregarding any individual perturbation label), not only achieves high performance in current metrics but actually outperforms most deep learning architectures without any learning taking place (see \ref{common-metrics-background} for examples).

Motivated by these troubling results, our goal in this work was to answer a simple question: ``\textit{why does the mean baseline perform well on standard perturbation response model metrics?}''. To answer this, we used simulated data as a discovery tool to understand the different dataset factors that can influence mean baseline performance and then confirmed our findings on two of the most commonly used benchmarking datasets \cite{norman2019exploring, replogle2022mapping}. We highlight the impact of systematic control bias (Fig.~\ref{fig:Graphical-abstract:a}) on inflating mean baseline performance and the impact of signal dilution (Fig.~\ref{fig:Graphical-abstract:b}) on contributing to mode collapse in model predictions. To address these issues, we introduce Weighted MSE (WMSE) and $R^2_w(\Delta)$, alternative metrics that are sensitive to differentially-expressed gene (DEG) signals against all perturbations (not control), and give null performance under mode collapse to the mean. We calibrate these metrics by introducing negative and positive baselines, including a novel technical duplicate baseline (Fig.~\ref{fig:Graphical-abstract:c}) which gives a realistic estimate of optimal performance given the intrinsic variance of the dataset. Moreover, we show that WMSE can be used as a training loss to prevent mode collapse and improve model performance (Fig.~\ref{fig:Graphical-abstract:c}). 

Together, these advances allow more transparent assessment of perturbation-response models and provide a general strategy for improving model performance. The code to replicate our results is available at \url{https://github.com/shiftbioscience/ICML2025\_pertmodel\_metrics}.


\section{Background}
\label{background}

\subsection{Single-cell Perturbation Data}

Single-cell RNA sequencing (scRNA-seq) measures the abundance of RNA transcripts in thousands-to-millions of individual cells, producing a gene-by-cell count matrix $\textbf{X}\in\mathbb{N}^{g\times n_T}$. Owing to low capture efficiency and drop-out, $\textbf{X}$ is sparse (typically $>90\%$ zeros), and is modeled well by a negative-binomial distribution. Standard pre-processing includes library-size normalization, log1p transformation, and feature selection of $\sim2$--$5$k most highly variable genes. 

Large-scale perturbation screens combine CRISPR or chemical interventions with scRNA-seq, yielding paired control–perturbation observations suitable for supervised learning. The screening technology Perturb-seq knocks out, represses (CRISPRi) or activates (CRISPRa) target genes prior to measurement of gene expression \cite{dixit2016perturb}. Perturb-seq datasets are often used to train perturbation-response models that aim to generalize to unseen perturbations or tissue types, with the hope of enabling large-scale \textit{in silico} screens where millions of perturbations can be tested without the need for costly and time-consuming wet lab experiments.

\subsection{Perturbation-Response Models}
Predictive models fall into four archetypes. (i) Simple linear baselines: ridge or principal-component regression that extrapolate additively from control and single-perturbation means. (ii) Autoencoder-based models: scGen, CPA and scVI fine-tune autoencoders to encode a cell and an intervention; counterfactuals are obtained by vector arithmetic in latent space \cite{lotfollahi2019scgen, lotfollahi2023predicting, lopez2018deep}; (iii) Prior knowledge graph learning: GEARS learns gene embeddings on a co-expression graph and perturbation embeddings on a gene-ontology graph, then decodes their interaction to predict expression shifts \cite{roohani2024predicting}. (iv) Transformer-based foundation models: scGPT \cite{cui2024scgpt} and scFoundation \cite{hao2024large} pre-train on millions of cells and adapt to perturbation tasks via conditioning tokens. Although every paradigm reflects key assumptions and inductive biases, they all strive to learn a conditional generation function $\hat{\mathbf{X}}^p = f(p, \mathbf{X^c})$ that predicts how the control cell population $\mathbf{X^c}$ would respond to perturbation $p$.

\begin{table*}[]
\centering
\caption{Parameter space used for simulation and real data experiments.}
\label{tab:parameters}
\begin{tabular}{@{}cllll@{}}
\toprule
\textbf{Parameter} & \textbf{Effect}                        & \textbf{Range Simulation} & \textbf{Range \textit{Norman19}} & \textbf{Range \textit{Replogle22}} \\ \midrule
$g$                & Number of genes in the dataset         & Linear: $1000-8192$       & $\log_2$: $2-8192$       & $\log_2$: $2-8192$         \\
$n_0$              & Number of control cells                & $\log$: $10-8192$         & $\log_2$: $1-8192$       & $\log_2$: $1-8192$         \\
$n_p$              & Number of cells per perturbation       & $\log$: $10-256$          & $\log_2$: $2-256$        & $\log_2$: $2-64$           \\
$k$                & Number of perturbations per dataset    & $\log$: $10-2000$         & $\log_2$: $1-175$        & $\log_2$: $1-1334$         \\
$\beta$            & Amount of systemic bias in the control & Linear: $0-2$             & Linear: $0-2$           & Linear: $0-2$             \\
$\delta$           & Probability of perturbing a gene       & Linear: $0.001-0.1$       & Quantiles: $0-1$        & Quantiles: $0-1$          \\
$\epsilon$         & Multiplicative effect of perturbations & $\log$: $1.2-5.0$         & Quantiles: $0-1$        & Quantiles: $0-1$          \\
$\mu_l$            & Library size scaling (Data quality)    & $\log$: $0.2-5.0$         & Deciles: $0-1$        & Deciles: $0-1$          \\ \bottomrule
\end{tabular}
\end{table*}

\subsection{Pitfalls of Common Performance Metrics}
\label{common-metrics-background}

Performance is generally quantified at the ``pseudobulk'' (aggregation of single cells) level after averaging all ground truth and predicted cell profiles per perturbation. This aggregation helps with the sparsity of scRNA-seq data and turns the task into an average effect prediction problem. Although numerous metrics have been introduced, the two most reported ones are MSE (or MAE) and Pearson($\Delta$). While MSE is defined as the average $L_2$ error, Pearson($\Delta$) aspires to capture perturbation effects with respect to a control mean profile $\mu^c$. For a single perturbation with average profile $\mu^p$ and predicted profile $\hat{\mu}^p$, Pearson($\Delta$) is defined as $r(\mu^p-\mu^c, \hat{\mu}^p-\mu^c) = r(\Delta^p, \hat{\Delta}^p)$, which aims to assess whether predicted changes from control are in the same direction as in real data.


Although intuitive, Pearson($\Delta$) faces two main limitations: (i) it does not consider the scale of change and hence predictions that do not capture the dynamic range of the true $\Delta^p$ signal can still perform well on this metric, and (ii) it heavily relies on the definition of a control population which might be biased depending on the dataset or the perturbation type. If that is the case, a mean baseline will perform quasi-optimally because the effect of having any perturbation dominates over the unique effect of a specific perturbation (Fig.~\ref{fig:Graphical-abstract:a}). On the other hand, the MSE suffers from signal dilution. As illustrated in Fig.~\ref{fig:Graphical-abstract:b}, MSE treats every error equally. Thus, it provides low error estimates for predictions that fail to capture important, low-dimensional biological changes but accurately track the general data distribution (e.g., the mean baseline). To address these problems, prior work has computed these metrics only on a subset of perturbation-specific DEGs, genes which change expression significantly as a result of a perturbation when compared to a control population. However, if used as an optimization approach, this requires sparse supervision which causes issues with genes not commonly recognized as DEGs (e.g., a gene only detected once as upregulated in a single training perturbation will systematically be learned as highly expressed). And second, this filtering makes performance metrics blind to true negatives (perturbations with no DEGs).

Multiple recent benchmarks that employ these and other metrics have independently found that predicting the mean baseline (the mean of all perturbed cells; $\mu^{all}$) often matches or surpasses state-of-the-art models. \citet{li2024benchmarking} assessed ten methods across multiple modeling tasks where the mean baseline achieved the lowest MAE($\Delta$) and nearly the highest Pearson($\Delta$). Analyzing four Perturb-seq datasets, \citet{csendes2025benchmarking} found that the mean baseline exceeds scGPT and scFoundation on Pearson($\Delta$). \citet{wenteler2024perteval} showed that the mean baseline tracks top-20 DEG effects as closely as scGPT, Geneformer, or UCE \cite{cui2024scgpt, theodoris2023transfer, rosen2023universal}. Finally, \citet{ahlmann2024deep} reported that four foundation and two deep-learning models fail to beat a mean baseline in evaluations across multiple datasets.


\section{Methods}
\label{methods}

\subsection{\textit{In silico} Simulations}
\label{in-silico-simulations}
We model synthetic datasets containing $n_0$ control cells and $k$ perturbations with a constant number of $n_p$ observed cells per perturbation. Each cell is represented by a random raw count vector $X\in\mathbb{R}^g$ where each component represents the observed expression of a single gene in that cell under a unique perturbation. Mathematically, we model the expression value $X^p_{i, j}$ of the $i^{th}$ gene of perturbation $p$ in the specific cell $j$ as a negative binomial with a fixed per-gene dispersion and variable mean:
\begin{align}
    X^p_{i,j} &\sim \mathrm{NB} \left(\mu^p_{i,j}, \theta_i\right)
\end{align}

Where $\theta_i$ is the fixed dispersion of gene $i$ and $\mu^p_{i,j}$ captures the simulated perturbation effects as follows:
\begin{align}
    \mbox{Perturbations:}&~~~~ \mu^p_{i,j} = l_j\alpha^p_{i}(\mu^c_i+\beta\lambda_i)\\
    \mbox{Control:}&~~~~\mu^c_{i,j} = l_j\mu^c_i
\end{align}

$\mu^c_i$ being the average of control expression, $\lambda_i$ a scalar symbolizing a realistic systematic bias between the control population and all other perturbations (only depends on the gene), $\beta$ a global dataset parameter controlling the severity to which $\lambda_i$ is applied (zero for a perfectly centered control), $\alpha^p_{i}$ a multiplicative effect on gene $i$ associated with perturbation $p$, and $l_j$ the library size component and affects every gene of the cell $j$ equally. Both $\alpha^p_{i}$ and $l_j$ are random variables by themselves distributed as shown:

\begin{align}
    \alpha^p_{i} &\sim \begin{cases} 
    1, & P = 1-\delta \\
    1/\epsilon, & P = \delta/2 \\
    \epsilon, & P = \delta/2 
\end{cases}\\[12pt]
l_j &\sim \text{LogNormal}(\mu_l, \sigma^2_l)
\end{align}

Here, $\delta$ represents the average probability of perturbing a gene, $\epsilon>1$ the strength of the effect, and $\mu_l, \sigma_l^2$ the mean and variance of the library size scaling factor. All $\delta, \epsilon, \mu_l, \sigma_l^2$ are constant for the entire dataset. Following reasonable priors, we define $\lambda_i=\mu^{all}_{i} - \mu^c_{i}$ as the difference between the average perturbed expression and the control expression in the \textit{Norman19} dataset which is also used to estimate $\theta_i, \mu_i^c,$ and $\sigma_l^2$. Given this setup, we perform random sampling to generate an array of synthetic datasets from the parameter space in Table \ref{tab:parameters}. Following standard processing, every synthetic dataset is library-size normalized to $10^4$ counts per cell and log1p transformed. We generated $10^4$ synthetic datasets and evaluated Pearson($\Delta$) and MSE on 4 gene sets: (1) all genes, (2) affected genes, which reflects the true simulated perturbed genes ($\alpha_i^p\neq1$), (3) observed DEGs vs control and (4) observed DEGs vs the rest of the perturbations.

\subsection{Real Data Experiments}

To evaluate the realism of our simulated results, we created analogous experiments in real-world data. We processed and analyzed two datasets commonly used in benchmarks: (1) \textit{Norman19}, a CRISPRa Perturb-seq dataset with genes activated alone or in combos of two \cite{norman2019exploring} and (2) \textit{Replogle22}, a genome-wide CRISPRi Perturb-seq dataset \cite{replogle2022mapping}. Datasets were randomly downsampled such that each perturbation label had the same number of cells ($256$ for \textit{Norman19} and $64$ for \textit{Replogle22}). For both datasets, we selected the top $8192$ highly-variable genes using the \texttt{highly\_variable\_genes} function from the scanpy package \cite{wolf2018scanpy}. We then used the \texttt{rank\_genes\_groups} function from scanpy with the \texttt{t-test\_overestim\_var} method to calculate DEGs with respect to the control cells (DEGs vs Control) and with respect to all other perturbations (DEGs vs Rest). A detailed description of the experiments performed on real data is provided in Appendix \ref{real-data-simulation} and high-level parameter ranges are available in Table \ref{tab:parameters}.  

\begin{table*}[ht]
\centering
\caption{Pearson correlations between metric performance of the mean baseline ($\hat{\mu}^p=\mu^{all}$) and dataset parameters in simulation experiments. Correlation values lower than $-0.2$ are highlighted in {\color[HTML]{3531FF} \textbf{blue}} and higher than $0.2$ are highlighted in {\color[HTML]{FE0000} \textbf{red}}.}
\label{tab:simulation-results}
\begin{tabular}{@{}cccccccccc@{}}
\toprule
Metric                          & Gene Group      & $\epsilon$                            & $g$                                   & $n_p$                                 & $\beta$                              & $\delta$                              & $n_0$                                 & $k$                                   & $\mu_l$ \\ \midrule
                                & All             & {\color[HTML]{3531FF} \textbf{-0.32}} & 0.00                                  & {\color[HTML]{FE0000} \textbf{0.28}}  & {\color[HTML]{FE0000} \textbf{0.54}} & {\color[HTML]{3531FF} \textbf{-0.28}} & {\color[HTML]{3531FF} \textbf{-0.48}} & -0.10                                 & 0.03    \\
                                & Affected        & {\color[HTML]{3531FF} \textbf{-0.25}} & -0.08                                 & 0.06                                  & 0.12                                 & -0.09                                 & -0.14                                 & {\color[HTML]{3531FF} \textbf{-0.68}} & -0.01   \\
                                & DEGs vs Control & {\color[HTML]{3531FF} \textbf{-0.50}} & -0.03                                 & 0.00                                  & {\color[HTML]{FE0000} \textbf{0.45}} & {\color[HTML]{3531FF} \textbf{-0.28}} & {\color[HTML]{3531FF} \textbf{-0.39}} & {\color[HTML]{3531FF} \textbf{-0.23}} & -0.13   \\
\multirow{-4}{*}{Pearson($\Delta$) $(\uparrow)$} & DEGs vs Rest    & {\color[HTML]{3531FF} \textbf{-0.33}} & -0.04                                 & {\color[HTML]{FE0000} \textbf{0.20}}  & {\color[HTML]{FE0000} \textbf{0.35}} & 0.00                                  & {\color[HTML]{3531FF} \textbf{-0.40}} & {\color[HTML]{3531FF} \textbf{-0.43}} & -0.03   \\ \midrule
                                & All             & {\color[HTML]{FE0000} \textbf{0.45}}  & {\color[HTML]{3531FF} \textbf{-0.44}} & {\color[HTML]{3531FF} \textbf{-0.35}} & 0.04                                 & {\color[HTML]{FE0000} \textbf{0.42}}  & -0.01                                 & 0.02                                  & -0.17   \\
                                & Affected        & {\color[HTML]{FE0000} \textbf{0.83}}  & {\color[HTML]{3531FF} \textbf{-0.42}} & -0.02                                 & 0.03                                 & 0.01                                  & 0.01                                  & 0.05                                  & -0.02   \\
                                & DEGs vs Control & {\color[HTML]{FE0000} \textbf{0.70}}  & {\color[HTML]{3531FF} \textbf{-0.46}} & {\color[HTML]{3531FF} \textbf{-0.25}} & 0.01                                 & -0.06                                 & 0.06                                  & 0.05                                  & -0.12   \\
\multirow{-4}{*}{MSE $(\downarrow)$}           & DEGs vs Rest    & {\color[HTML]{FE0000} \textbf{0.67}}  & {\color[HTML]{3531FF} \textbf{-0.48}} & {\color[HTML]{3531FF} \textbf{-0.26}} & 0.03                                 & -0.08                                 & 0.01                                  & 0.04                                  & -0.14   \\ \bottomrule
\end{tabular}
\end{table*}

\subsection{Proposed Metrics}
\label{proposed-metrics-methods}

\subsubsection{Weighted Delta $R^2$: $R^2_w(\Delta)$}
Given a set of positive weights $\{w_i\}$ that add to one, average perturbed expression levels $\{\mu_i^{all}\}$, ground truth expression levels $\{\mu^p_i\}$ and pseudobulked predicted values $\{\hat{\mu}^p_i\}$, $i \in \{1, 2, \dots, g\}$ with $g$ the number of genes in the dataset, we define $R^2_w(\Delta)$ for a single perturbation as follows:
\begin{align}
    R^2_w(\Delta) &= 1 - \frac{\sum_i w_i(\Delta_i-\hat{\Delta}_i)^2}{\sum_i w_i(\Delta_i-\bar{\Delta}_w)^2}\\
    \bar{\Delta}_w &= \sum_{i=1}^{g} w_i \Delta_i
\end{align}

Where $\Delta_i = \mu^p_i - \mu_i^{all}$ and $\hat{\Delta}_i = \hat{\mu}^p_i - \mu_i^{all}$ represent the real and predicted changes from the average of all perturbed cells respectively.  Note that reference values for delta computation $\mu_i^{all}$ are the center of all perturbed cells in the dataset instead of the traditional definition which computes against the control population $\mu_i^c$. We propose $R^2_w(\Delta)$ as a significantly more stringent alternative to Pearson($\Delta$) with the following four advantages. (i) As a goodness of fit metric, the scale and dynamic range of the predictions does matter. It is not enough to estimate the direction of change as with Pearson($\Delta$). (ii) Because we set the reference to the mean of all perturbed cells in the dataset, there is, by definition, no systematic bias that can inflate metrics unintentionally. (iii) Because of the properties of $R^2$, any constant average predictions $(\hat{\mu}^p=\mu^{all})$ will yield a strictly negative result for any specific perturbation (see Appendix \ref{upper-bound-r2} for derivation). (iv) This metric, while still computing in full transcriptomic space, can prioritize more biologically significant genes like DEGs by changing the weights definition.

\subsubsection{Weighted Error: WMSE}
\label{wmse-def}

We propose to evaluate the perturbation prediction task with a modified version of the classical MSE regression metric defined as follows for a single perturbation:
\begin{align}
    \text{WMSE} &= \sum_{i=1}^{g}w_i(\mu^p_i-\hat{\mu}^p_i)^2
\end{align}

While simple, this modification of regular MSE addresses the main pitfalls of error metrics currently used in the task (see \ref{common-metrics-background}). Unlike standard MSE, WMSE allows for gene signal prioritization, such as for perturbation-specific DEGs. In other words, WMSE is more sensitive to perturbation-specific signals, which is particularly important given that only a small proportion of genes change meaningfully in response to each perturbation \cite{nadig2025transcriptome}. Moreover, WMSE can directly replace MSE as training loss for many models allowing for biologically meaningful supervision.

\subsubsection{Weights Definition}

Although the weight set $\{w_1, w_2, \dots, w_g\}$ may be arbitrarily chosen to highlight any biologically relevant signal on the data, here we choose the weights to prioritize perturbation-specific DEGs. This intuitively assigns higher importance to genes that change meaningfully in a perturbation while still allowing consideration of the whole transcriptome. Our weight computation procedure for a single perturbation is the following: (i) we determine $t-$scores for every gene with respect to the rest of the perturbed cells in the dataset (using scanpy's \texttt{sc.tl.ran\_genes\_groups} \cite{wolf2018scanpy} function with \texttt{method='t-test\_overestim\_var'} and \texttt{reference='rest'}) (ii) we apply an absolute value transformation, (iii) we perform min-max normalization to the $[0, 1]$ range, (iv) we square the weights to accentuate differences, and (v) we normalize the whole weight set to add up to $1$. 
The key differentiator of this method is the use of all other perturbed cells as reference for DEG calculation (DEGs vs Rest) instead of the experiment's control (DEGs vs Control). This selection ensures the prioritized genes are the ones that make that perturbation unique from all the others without introducing control bias.

\begin{figure*}[ht]
\vskip 0.2in
\begin{center}
\centerline{\includegraphics[width=\textwidth]{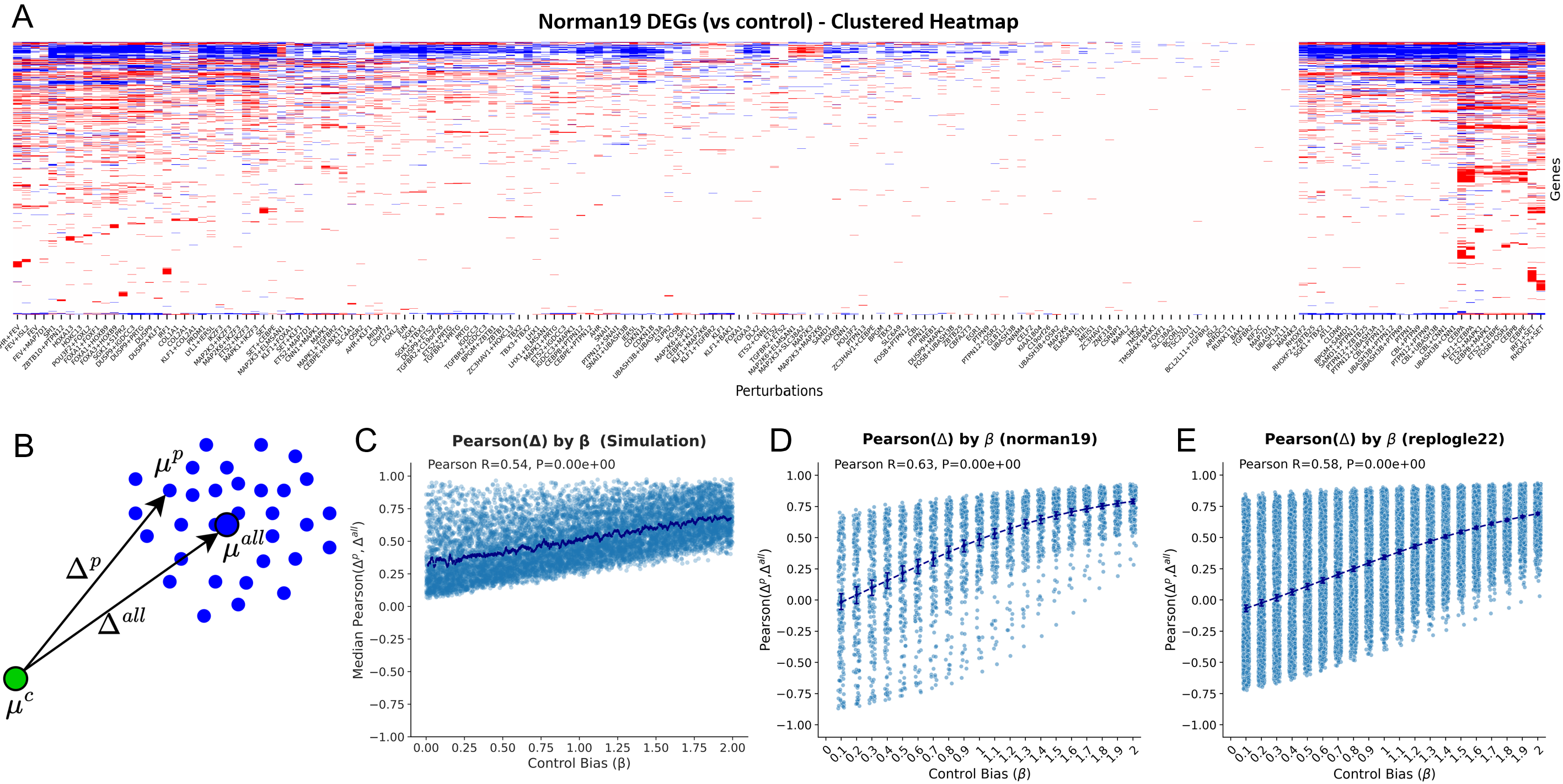}}
\caption{(A) Trinary (up, down, or unchanged) clustermap of significant DEGs for each perturbation compared to control. (B) Schematic showing high correlation between $\Delta^p$ and $\Delta^{all}$ due to the systematic bias of $\mu^c$. (C) Pearson($\Delta$) performance of the mean baseline ($\mu^{all}$) under increasing control bias ($\beta$) in simulations. Trend line shows moving average. (D) Pearson($\Delta$) performance of $\mu^{all}$ under increasing $\beta$ in the \textit{Norman19} dataset ($\beta=0$ no control bias, $\beta=1$ exact dataset control bias, $\beta=2$ double the dataset control bias). Trend line shows mean and 95\% CI of the mean. (E) Same as (D) in the \textit{Replogle22} dataset.}
\phantomsubcaption\label{control-bias-fig:a}
\phantomsubcaption\label{control-bias-fig:b}
\phantomsubcaption\label{control-bias-fig:c}
\phantomsubcaption\label{control-bias-fig:d}
\phantomsubcaption\label{control-bias-fig:e}
\label{control-bias-fig}
\end{center}
\vskip -0.3in
\end{figure*}

\subsection{Technical Duplicate Baseline}
\label{tech-dup-method}

scRNA-seq perturbation data poses multiple challenges for reproducible perturbation modeling, such as low capture, difficult annotation of perturbed cells, low perturbation efficiency, and high dimensionality which all contribute to high variance in average effect estimation for any given perturbation. To address this problem, we propose a technical duplicate baseline which tries to answer a simple question: ``\textit{how would a technical duplicate of the dataset perform in predicting a mean perturbation effect?}''. Achieving this level of performance for a model would mean that its prediction errors are comparable to the variance of the experiment itself, defining a performance ceiling. We compute this baseline by randomly dividing the population of cells receiving a perturbation in half and using one half of the cells to predict the other half.

\subsection{MSE vs WMSE Training}

To assess whether WMSE can serve not only as an evaluation metric but also as a useful learning objective, we retrained GEARS from scratch with default hyperparameters under three loss functions: (i) its original loss, (ii) the standard (unweighted) MSE, and (iii) WMSE. For \textit{Norman19} we train GEARS for the combination prediction task using all single perturbations and half of the combination perturbations for training and validation with the remaining half of combinations for testing. For \textit{Replogle22} we perform the unseen gene prediction task where half the data is used for training and validation and the remaining perturbations are used for testing. In both cases, we train two GEARS models to get predictions for the all combinations or all unseen genes.


\section{Experimental Results}

\subsection{\textit{In silico} Screen}

We ran an \textit{in silico} screen as a discovery tool to pinpoint the factors that inflate mean-baseline performance (results summarized in Table \ref{tab:simulation-results}). As expected, both perturbation effect size ($\epsilon$) and gene perturbation probability ($\delta$) generally decrease baseline performance. This is expected as stronger perturbations give more distinct signals and thus $\mu^{all}$ (mean of all perturbed cells) is a worse approximation of any individual $\mu^p$ (mean of specific perturbation). Similarly, the number of genes $g$ reduced all error metrics due to library size normalization (more genes imply generally lower values post normalizing to $10^4$ counts). And finally, the number of cells per perturbation $n_p$ also increased performance, likely due to less sparsity in every pseudobulk $\mu^p$ estimation.


\begin{figure}[ht!]
\begin{center}
\centerline{\includegraphics[width=0.9\columnwidth]{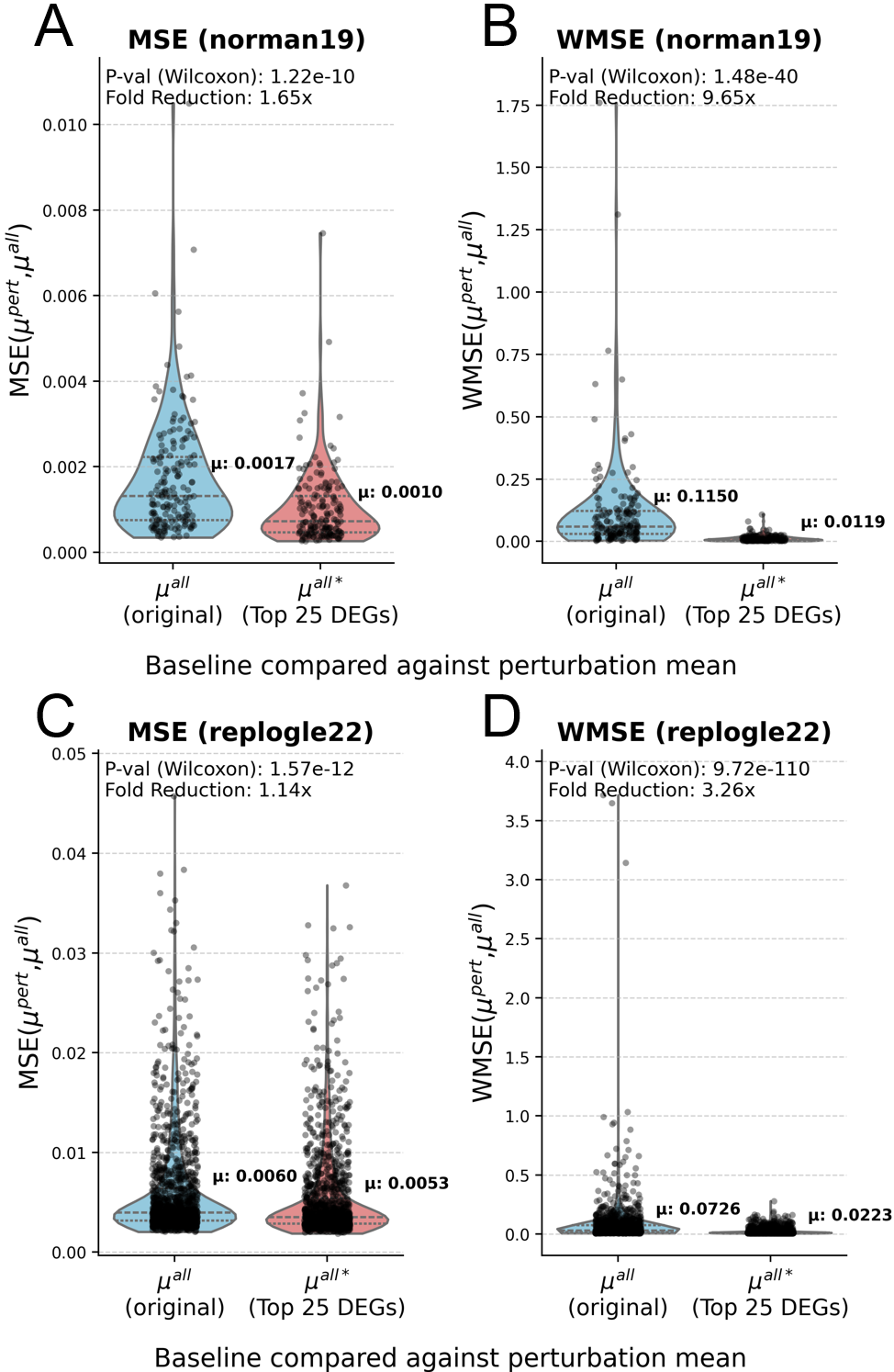}}
\caption{Weighted MSE is sensitive to differences in niche perturbation signals. (A) Violin plot showing MSE between true perturbation profiles $\mu^p$ and two possible predictions: the mean baseline $\mu^{all}$ and a modified version $\mu^{all*}$ which sets the top 25 DEGs to be perfectly predicted. (B) Same as (A) but MSE was weighted by the normalized DEG score (WMSE). (C-D) Same as (A-B) but for the \textit{Replogle22} dataset.}
\label{wmse-main-plot}
\phantomsubcaption\label{wmse-main-plot:a}
\phantomsubcaption\label{wmse-main-plot:b}
\phantomsubcaption\label{wmse-main-plot:c}
\phantomsubcaption\label{wmse-main-plot:d}
\end{center}
\vskip -0.4in
\end{figure}

Apart from these effects, three parameters strongly modulated Pearson($\Delta$): the magnitude of control bias ($\beta$), the number of control cells ($n_{0}$), and the total number of perturbations ($k$). An in-depth analysis of control bias influence is provided in the following while additional results concerning $n_{0}$ and $k$ are provided in Appendices \ref{n0-experiments} and \ref{k-experiments} respectively.

\subsection{Control Bias $\beta$}

Systematic bias is readily apparent when evaluating DEGs against control. This is illustrated in the \textit{Norman19} dataset in which a substantial proportion of DEGs are shared across multiple perturbations (Fig.~\ref{control-bias-fig:a} and Supplemental Figs. \ref{figs1:b},\ref{figs1:c},\ref{figs1:d},\ref{figs1:e}). Notably, the \textit{Replogle22} dataset has fewer cells per perturbation (64 compared to 256 in \textit{Norman19}), and thus less power to detect DEGs. Yet, we still observe that many DEGs are conserved across perturbations in this dataset (Supplemental Fig.~\ref{figs1:a}). As shown in Fig.~\ref{control-bias-fig:b}, we hypothesized that the high performance of the mean baseline on Pearson($\Delta$) could be explained if the effect of any perturbation (compared to the control mean) is similar to the effect of all perturbations compared to control. Under this hypothesis, the global expression difference ($\Delta^{all}=\mu^{all}-\mu^c$) becomes highly correlated with perturbation-specific differences ($\Delta^p=\mu^{p}-\mu^c$) as the distinction between perturbed and control states dominates the more subtle differences among perturbations. In our simulations, control bias ($\beta$) showed a high correlation with mean baseline ($\mu^{all}$) performance on Pearson($\Delta$) (Fig.~\ref{control-bias-fig:c}, $r=0.54$). This behavior was also observed in real data when introducing or removing control bias, with even stronger correlations of $0.63$ and $0.58$ in the \textit{Norman19} (Fig.~\ref{control-bias-fig:d}) and \textit{Replogle22} (Fig.~\ref{control-bias-fig:e}) datasets, respectively. These findings suggest that even a modest control bias leads to increased mean baseline performance on the Pearson($\Delta$) metric without any actual learning taking place. Moreover, the \textit{Norman19} and \textit{Replogle22} datasets are not considered poor-quality and, indeed, are used widely for model benchmarking \cite{wu2024perturbench,li2024benchmarking,csendes2025benchmarking,li2024systematic,bendidi2024benchmarking,wenteler2024perteval,ahlmann2024deep}. Addressing control bias is likely not a matter of picking better datasets but rather picking a better reference. Instead of using non-targeting (NT) control cells, a more unbiased analysis might leverage all perturbed cells as the reference for DEG analysis and $\Delta$-based performance metrics. 

\subsection{DEG Score-weighted MSE (WMSE)}

While many perturbations strongly impact the transcriptome, leading to a large number of DEGs, most perturbations do not (Supplemental Fig.~\ref{figs1:e}). To address the problem of DEG signal dilution in perturbation modeling, we propose an MSE weighted by the strength of perturbation-specific DEGs (calculated with respect to all other perturbations; ``vs Rest'') (see \ref{wmse-def}). To evaluate the sensitivity of MSE and WMSE to niche DEG signals, we computed both metrics under two scenarios: for each perturbation we compared the perturbation mean ($\mu^p$) with (1) the mean of all perturbed cells ($\mu^{all}$), or (2) $\mu^{all*}$ in which we artificially set the gene expression of $\mu^{all}$ to be identical to $\mu^p$ for the top $25$ perturbation-specific DEGs (approximately $0.3\%$ of all genes). Despite the small modification, it alters the most important perturbation-specific signals. As expected, WMSE was significantly more sensitive to the niche DEG signals in $\mu^{all*}$ in both datasets compared with MSE (Fig.~\ref{wmse-main-plot}). Overall, these findings highlight the utility of DEG-aware metrics to capture perturbation-specific signals.

\begin{figure}[ht!]
\begin{center}
\centerline{\includegraphics[width=0.9\columnwidth]{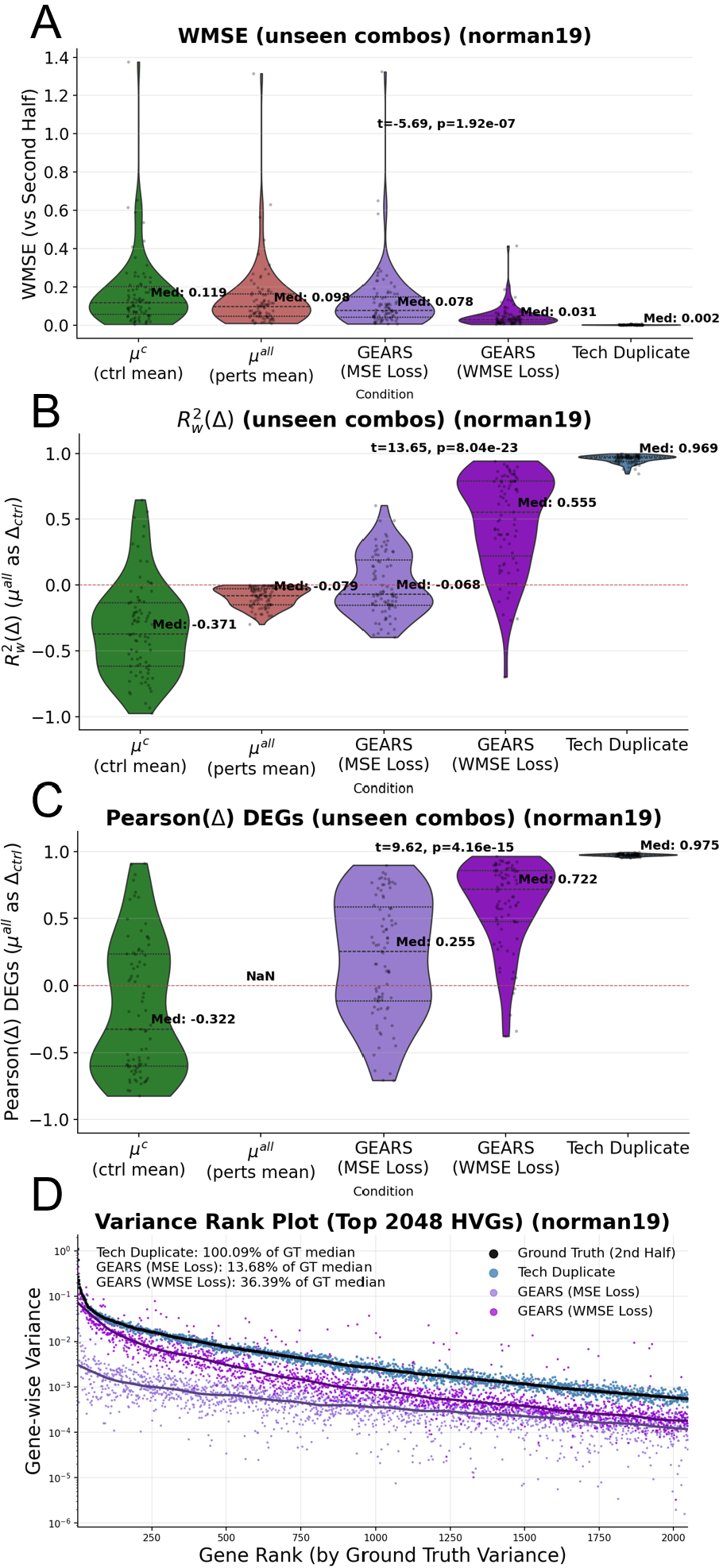}}
\caption{DEG score-weighted loss reduces mode collapse and improves model performance (\textit{Norman19}). (A) WMSE between prediction and ground-truth perturbation mean. X labels: $\mu^c$ (control mean), $\mu^{all}$ (mean of all perturbed cells), predictions from GEARS model with MSE or WMSE loss, and technical duplicate baseline. Means between GEARS MSE/WMSE compared with paired t-test. (B) Same as (A) but for $R^2_w(\hat{\Delta}^p,\Delta^{p})$, the DEG score-weighted $R^2$ between predicted ($\hat{\Delta}^p$) vs ground-truth perturbation effect ($\Delta^{p}$). For $\Delta$ calculations, $\mu^{all}$ is the reference. (C) Same as (B) but with Pearson correlation and filtering to only include perturbation-specific DEGs (vs Rest). (D) Plot showing the top 2048 highly-variable genes ranked by variance in the ground truth pseudobulked dataset.
}
\label{wmse-improvement-norman}

\phantomsubcaption\label{wmse-improvement-norman:a}
\phantomsubcaption\label{wmse-improvement-norman:b}
\phantomsubcaption\label{wmse-improvement-norman:c}
\phantomsubcaption\label{wmse-improvement-norman:d}
\end{center}
\vskip -0.4in
\end{figure}

\subsection{Improved Metrics with Scale Calibration}
\label{calibration}

As discussed above (see \ref{common-metrics-background}), common perturbation modeling metrics have multiple drawbacks: (1) MSE/MAE dilutes DEG signals (and filtering to top-$k$ DEGs requires arbitrary cutoffs and cannot measure non-DEG errors), and (2) Pearson($\Delta$) is sensitive to control bias and cannot measure whether predicted effects have a similar dynamic range to true perturbation effects. The result is that current metrics often cannot detect mode collapse of perturbation models (i.e., they predict $\mu^{all}$ regardless of perturbation label). To address these limitations we introduce two metrics (see \ref{proposed-metrics-methods}): WMSE and DEG score-weighted $R^{2}(\Delta)$ ($R_w^{2}(\Delta)$) in which $\mu^{all}$ is the $\Delta$ reference. Moreover, current benchmarking metrics often lack scale calibration, as negative and positive performance baselines are often not included. To address this limitation, we implement $\mu^c$ and $\mu^{all}$ as biased and uninformative negative baselines, respectively, and introduce a technical duplicate baseline that simulates ideal model performance (see \ref{tech-dup-method}).

As shown in Supplemental Fig.~\ref{figs3}, our baselines reveal that DEG-based weighting greatly expands the dynamic range of each metric and reinstates a coherent ranking: the biased control prediction performs worst, the mean baseline has null performance, and the technical duplicate baseline establishes the upper bound. Without weighting, this order collapses; in \textit{Replogle22}, for instance, the technical duplicate performs worst on both MSE and $R^{2}(\Delta)$ despite being the most informative predictor. The apparent anomaly highlights the signal-dilution problem.  Technical duplicate estimates are derived from only half the cells in each perturbation (32 in \textit{Replogle22}) and therefore carry more sampling noise than the population means returned by the baselines. Metrics that treat all genes uniformly reward this noise reduction, even though it is biologically irrelevant. Once genes are re-weighted by their perturbation-specific DEG statistics (WMSE, $R_w^{2}(\Delta)$), the technical duplicate’s superior capture of genuine signal outweighs its higher variance, whereas models that regress toward the dataset mean lose ground. Indeed, for a weak perturbation such as MRPL23, unweighted $\Delta$ vectors from the two technical duplicate halves show virtually no correlation, yet restricting the comparison to the five true DEGs restores a strong correspondence, precisely the behavior our weighting scheme is designed to reward (Supplemental Fig.~\ref{scatterplot-r2}).


\subsection{MSE vs WMSE Training}

With our calibrated metrics, we evaluated the performance of a well-established perturbation model, GEARS (Supplemental Fig.~\ref{figs3}). We found that GEARS displays decent performance in the combo-prediction task (\textit{Norman19}), which requires extrapolating from single-gene effects to two-gene combos. However, GEARS struggles to outperform the $\mu^{all}$ baseline in the more difficult unseen gene prediction task (\textit{Replogle22}), as it involves zero-shot predictions in a noisier dataset (Supplemental Fig.~\ref{figs3}). We hypothesized that, especially for the unseen gene task, GEARS may be experiencing mode collapse during training due to its unweighted MSE loss. To test this, we retrained GEARS with either MSE or WMSE loss on both datasets and evaluated its performance (Fig.~\ref{wmse-improvement-norman}, Supplemental Fig.~\ref{figs4}).

On \textit{Norman19}, the null ($\mu^{all}$) baseline yields strictly negative \(R^{2}_{w}(\Delta)\) values that cluster near zero; thus any positive score indicates genuine learning.  Switching the training objective from MSE to WMSE reduces the test-set WMSE by a factor of 2.5 and lifts the median \(R^{2}_{w}(\Delta)\) from \(-0.068\) to \(0.555\), bringing GEARS to within striking distance of the technical duplicate baseline. The same trend appears in Pearson($\Delta$) computed after filtering to DEGs (vs Rest) (Fig.~\ref{wmse-improvement-norman:c}), a metric whose definition contains no weighting and thus rules out information leakage from the training weights.  Gene-wise variance profiles (Fig.~\ref{wmse-improvement-norman:d}) reveal the likely source of this performance gain: WMSE penalizes mode collapse, so predictions recover much of the true perturbation variance instead of shrinking toward the dataset mean.

In the more difficult unseen gene prediction task (\textit{Replogle22}), WMSE again outperforms MSE on every weighted metric (though this improvement is smaller compared to the combo task) (Supplemental Fig.~\ref{figs4}). When evaluation is restricted to the top 5\% of perturbations ranked by DEG count (Supplemental Fig.~\ref{figs6}), WMSE-trained GEARS overtakes the baseline, demonstrating that the model can capture perturbation-specific effects once the biological signal is strong enough. Moreover, we found that gene-wise variances align far more closely with ground truth when WMSE is used (Supplemental Fig.~\ref{figs4:d}), suggesting again that performance gains likely result from amelioration of mode collapse.

Taken together, these results indicate that DEG-based weighting steers optimization towards sparse, high-variance predictions that better reflect real perturbation effects, while simultaneously aligning with an intuitive calibration in $R^{2}_{w}(\Delta)$: null predictors cluster near zero, ordinary learning objectives improve modestly, and WMSE moves performance toward the empirical ceiling set by technical duplicate baseline. The unseen combo task (\textit{Norman19}) is an easier task because information about all single genes is provided to models during training, and thus there is some data leakage when predicting unseen gene combo effects at test time. While this may inflate our estimation of WMSE-driven gains on that task, the parallel improvement in the \textit{Replogle22} zero-shot setting confirms that weighting does provide a genuine generalization advantage.  Embedding this inductive bias more explicitly in future architectures should help models remain competitive even when perturbations are weak or data quality is suboptimal.


\section{Conclusion}

Recent benchmarks reveal that predicting the perturbed dataset mean often performs much better than expected without any learning taking place and often surpasses fitted models' performance on common evaluation metrics. From our analyses on \textit{in silico} and real-world datasets, we traced this behavior to bias in control cells and metric artifacts that reward mode collapse. Our conclusions produced a four-step remedy: (i) use the mean of all perturbed cells to remove systematic control bias in $\Delta$ and DEG calculations; (ii) adopt DEG-score weighted metrics ($\Delta R^{2}_{w}$, WMSE) that penalize mode collapse while retaining transcriptome-wide coverage; (iii) calibrate all metrics with negative ($\mu^c$), null ($\mu^{all}$), and positive (technical duplicate) baselines; and (iv) implement DEG-aware optimization objectives (e.g., WMSE). Under this protocol, the mean baseline falls to null performance and models that capture perturbation-specific effects rise to the top.

\section*{Broader Impact}

Accurate \textit{in silico} perturbation response models can shorten drug‑discovery cycles, cut laboratory costs, and reduce animal use by flagging promising candidates before any wet‑lab work.  Metrics that reward degenerate averages, however, risk elevating brittle models that provide uninformative predictions. By reducing reference bias, implementing calibrated, DEG-aware metrics, and introducing an optimization approach that penalizes mode collapse, our work enables scientists to avoid misleading metric artifacts and steer their resources toward building and evaluating better perturbation response models.


\bibliography{references}
\bibliographystyle{icml2025}


\newpage
\appendix
\onecolumn

\section{Simulation on Real Data}
\label{real-data-simulation}

For every parameter in the simulated data, we designed real data experiments as follows:
\begin{itemize}
    \item $g$: In a $\log_2$ sequence from 2-8192, we randomly downsampled the data to $N$ genes.
    \item $n_0$: In a $\log_2$ sequence from 1-8192, we randomly downsampled the control cell population to $N$ cells.
    \item $n_p$: In a $\log_2$ sequence from 2-256 (for \textit{Norman19}) and 2-64 (for \textit{Replogle22}), we randomly selected $N$ cells for each perturbation label.
    \item $k$:  In a $\log_2$ sequence from 1-175 (for \textit{Norman19}) and 1-1334 (for \textit{Replogle22}), we randomly selected $N$ perturbations. We repeated downsampling with 10 random seeds.
    \item $\beta$: We calculated the $\Delta$ between the mean of all perturbed cells ($\mu^{all}$) and the mean of the control cells ($\mu^c$). We then created synthetic control data by interpolating in equivalent steps of $0.1\Delta$ between $\mu^{all}$ ($0\Delta$) and $\mu^c$ ($1\Delta$), terminating the interpolation at $2\Delta$.
    \item $\delta$: Because there was no real-data equivalent of this simulated parameter, we mimicked it by downsampling the data to include perturbations with variable numbers of detected DEGs. We first ranked perturbations by the number of significant DEGs detected. We then downsampled the data by selecting perturbations in $20\%$ quantile windows of normalized ranks ($0-0.2$, $0.1-0.3$, ...) such that $0-0.2$ had the weakest perturbations and $0.8-1.0$ had the strongest.
    \item $\epsilon$: To evaluate perturbation strength, we downsampled each dataset via the following procedure. For each perturbation, we ranked all genes by their absolute t-test metric score (obtained during DEG calculations). Ranks were binned into $10\%$ quantiles. Data were downsampled to generate one dataset per decile such that each perturbation only contained the genes within the relevant DEG quantile. Thus, for quantile $0-0.1$, the data contained only the least differentially expressed genes within each perturbation, and for $0.9-1.0$ the data contained the most.
    \item $\mu_l$: Within each perturbation, cells were ranked by library size, and the ranks were binned into deciles. Data were downsampled by selecting only cells belonging to each decile in sequence ($0-0.1$, $0.1-0.2$, ...).
\end{itemize}

\begin{table}[ht]
\centering
\caption{Pearson correlations between metrics and simulation parameters. Comparison between real data experiments and simulation experiments for all the genes. Correlation values lower than $-0.2$ are highlighted in {\color[HTML]{3531FF} \textbf{blue}} and Correlation values higher than $0.2$ are highlighted in {\color[HTML]{FE0000} \textbf{red}}.}
\label{tab:simulation-results-aggregated}
\begin{tabular}{@{}cccccccccc@{}}
\toprule
\textbf{Metric}                 & \textbf{Dataset}               & \textbf{$\epsilon$}                   & \textbf{$g$}                          & \textbf{$n_p$}                        & \textbf{$\beta$}                     & \textbf{$\delta$}                     & \textbf{$n_0$}                        & \textbf{$k$} & \textbf{$\mu_l$}                      \\ \midrule
                                & Norman19                       & {\color[HTML]{333333} \textbf{-}}     & {\color[HTML]{FE0000} \textbf{0.36}}  & {\color[HTML]{FE0000} \textbf{0.39}}  & {\color[HTML]{FE0000} \textbf{0.63}} & -0.19                                 & {\color[HTML]{3531FF} \textbf{-0.64}} & -0.09        & -0.09                                 \\
                                & Replogle22                     & {\color[HTML]{333333} \textbf{-}}     & 0.10                                  & {\color[HTML]{FE0000} \textbf{0.27}}  & {\color[HTML]{FE0000} \textbf{0.58}} & -0.06                                 & {\color[HTML]{3531FF} \textbf{-0.74}} & -0.03        & 0.07                                  \\
\multirow{-3}{*}{Pearson Delta} & \multicolumn{1}{l}{Simulation} & {\color[HTML]{3531FF} \textbf{-0.32}} & 0.00                                  & {\color[HTML]{FE0000} \textbf{0.28}}  & {\color[HTML]{FE0000} \textbf{0.54}} & {\color[HTML]{3531FF} \textbf{-0.28}} & {\color[HTML]{3531FF} \textbf{-0.48}} & -0.10        & 0.03                                  \\ \midrule
                                & Norman19                       & -                                     & {\color[HTML]{FE0000} \textbf{0.21}}  & {\color[HTML]{3531FF} \textbf{-0.74}} & -                                    & {\color[HTML]{FE0000} \textbf{0.70}}  & -                                     & -            & {\color[HTML]{3531FF} \textbf{-0.42}} \\
                                & Replogle22                     & -                                     & 0.17                                  & {\color[HTML]{3531FF} \textbf{-0.85}} & -                                    & {\color[HTML]{FE0000} \textbf{0.63}}  & -                                     & -            & {\color[HTML]{3531FF} \textbf{-0.58}} \\
\multirow{-3}{*}{MSE}           & \multicolumn{1}{l}{Simulation} & {\color[HTML]{FE0000} \textbf{0.45}}  & {\color[HTML]{3531FF} \textbf{-0.44}} & {\color[HTML]{3531FF} \textbf{-0.35}} & 0.04                                 & {\color[HTML]{FE0000} \textbf{0.42}}  & -0.01                                 & 0.02         & -0.17                                 \\ \bottomrule
\end{tabular}
\end{table}

\section{Upper Bounds of $R^2_w(\Delta)$ Under Constant $\hat{\mu}^p=\mu^{all}$ Predictions}
\label{upper-bound-r2}

Given the definition of the metric:
\begin{align}
    R^2_w(\Delta) &= 1 - \frac{\sum_i w_i(\Delta_i-\hat{\Delta}_i)^2}{\sum_i w_i(\Delta_i-\bar{\Delta}_w)^2}\\
    \bar{\Delta}_w &= \sum_{i=1}^{g} w_i \Delta_i\\
    \Delta_i &= \mu^p_i - \mu_i^{all}\\
    \hat{\Delta}_i &= \hat{\mu}^p_i - \mu_i^{all}
\end{align}

Under a constant prediction $\hat{\mu}^p_i = \mu_i^{all}$ any predicted delta becomes $\hat{\Delta}_i=0$ and the overall metric reduces to:

\begin{align}
    R^2_w(\Delta) &= 1 - \frac{\sum_i w_i\Delta_i^2}{\sum_i w_i(\Delta_i-\bar{\Delta}_w)^2}\\
\end{align}

Which can be easily shown to be negative by expanding $\sum_i w_i(\Delta_i-\bar{\Delta}_w)^2$ as follows:

\begin{align}
    \sum_i w_i(\Delta_i-\bar{\Delta}_w)^2 &= \sum_i w_i\Delta^2_i - 2\bar{\Delta}_w\sum_iw_i\Delta_i + \bar{\Delta}^2_w\sum_iw_i 
\end{align}

Because the set $\{w_i\}$ is normalized to add up to one and under the definition of $\bar{\Delta}_w$ the expression can be reduced to:

\begin{align}
    \sum_i w_i(\Delta_i-\bar{\Delta}_w)^2 &= \sum_i w_i\Delta^2_i - 2\bar{\Delta}_w^2 + \bar{\Delta}^2_w\\
    &= \sum_i w_i\Delta^2_i - \bar{\Delta}_w^2
\end{align}

Then, rewriting the original metric value under the expansion we get:

\begin{align}
    R^2_w(\Delta) &= 1 - \frac{\sum_i w_i\Delta_i^2}{\sum_i w_i\Delta^2_i - \bar{\Delta}_w^2}\\
\end{align}

From which the rightmost fraction is clearly bounded to be positive (the original fraction was between squared quantities) and greater or equal than $1$ making $R^2_w(\Delta) \leq 0$ under the constant prediction case for any perturbation.

\section{Control Bias and Number of Control Cells ($n_0$)}
\label{n0-experiments}

\begin{itemize}
    \item Having observed this control bias, we questioned whether better sampling of the control population might be sufficient to reduce it by better approximating the center of the data (Supplemental Fig.~\ref{figs1:f}). 
    \item We simulated increasingly higher numbers of control cells and found this reduced the predicting accuracy of the dataset mean (indicating a most robust control) (Supplemental Fig.~\ref{figs1:g}). However, the simulation also demonstrated that there are diminishing returns from continuing to sample the control population beyond 1000 cells. A similar effect was observed in both the \textit{Norman19} and \textit{Replogle22} dataset (Supplemental Fig.~\ref{figs1:h} and Supplemental Fig.~\ref{figs1:i}). These results highlight that, while greater sampling of the control cell population is sufficient to reduce bias, it cannot eliminate it. Thus, metrics which hinge upon the presence of an unbiased control cell population are fundamentally confounded by these effects. This poses a particular challenge for metrics based on deltas (such as the Pearson($\Delta$)) and DEGs, when deltas and DEGs are calculated with respect to the control cell population. This is a common practice in the field today, as evidenced by the widespread use of these metrics in recent papers \cite{gong2023xtrimogene,istrate2024scgenept,roohani2024predicting,cui2024scgpt,li2024benchmarking,wenteler2024perteval,tang2024scperb,csendes2025benchmarking}. 
\end{itemize}

\section{Number of perturbations ($k$)}
\label{k-experiments}
\begin{figure}[ht]
\vskip 0.2in
\begin{center}
\centerline{\includegraphics[width=\textwidth]{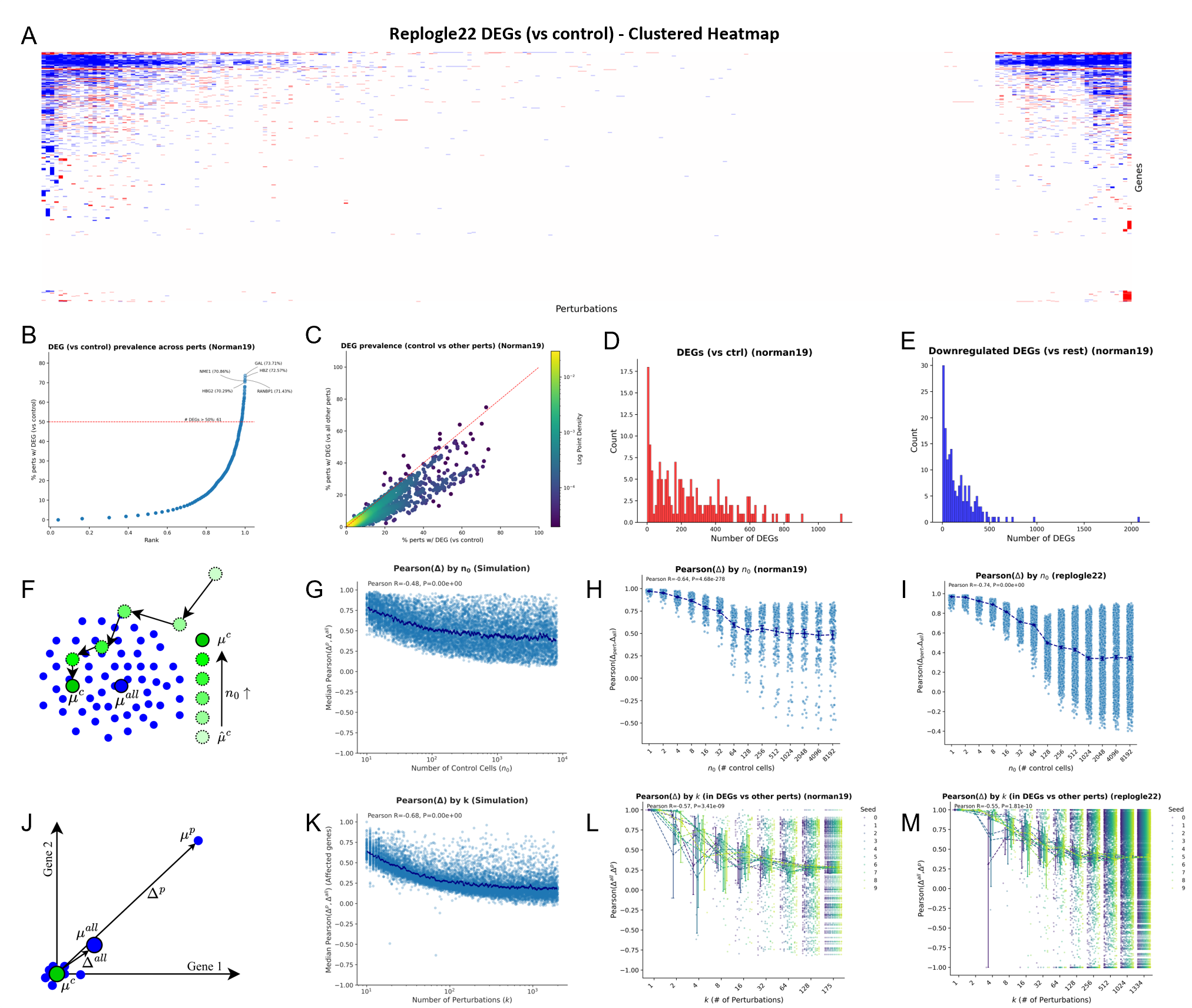}}
\caption{Supplemental figure accompanying Figure 2. (A) Trinary (up, down, or unchanged) clustermap of significant differentially expressed genes of every perturbation against the control population for the \textit{Replogle22} dataset. Genes and perturbations were downsampled randomly (to 256 and 2048 respectively) due to restrictions in plotting software. (B) Rank plot showing the percentage of perturbations detecting each DEG (vs Control) in the \textit{Norman19} dataset. Annotation shows the top shared genes and the number of genes shared by most perturbations (61).  (C) Scatter plot (with density histogram) showing the percentage of perturbations detecting each DEG (vs Control) compared with the percentage of perturbations detecting each DEG (vs all other perturbations). (D) Histogram showing distribution in terms of number of significant DEGs per perturbation in the \textit{Norman19} dataset. (E) Same as (D) but for DEGs calculated with respect to all other perturbations (vs Rest). (F) Diagram showing the effect of increasing the number of control cells ($n_0$) on improving estimation of the control mean ($\mu^c$) also reducing systematic bias. (G) Plot showing the effect of increasing control cell number ($n_0$) on Pearson($\Delta^p$,$\Delta^{all}$), which is the similarity between $\mu^p - \mu^{c}$ and $\mu^{all} - \mu^{c}$, in simulated data. (H-I) Same as (G), but in real datasets \textit{Norman19} and \textit{Replogle22} respectively. (J) Plot illustrating the biasing of $\mu^{all}$ by a strong perturbation $\mu^p$. The effect of this bias is to increase the similarity in direction between $\Delta^p$ and $\Delta^{all}$, especially when the dataset contains fewer perturbations in the first place to moderate this single-perturbation influence. (K) Plot showing the effect of number of perturbations in a simulated dataset ($k$) on the Pearson($\Delta^p$,$\Delta^{all}$). (L-M), same as (K) except in real datasets and with 10 random seeds for selection of $k$ perturbations.
}
\label{figs1}
\phantomsubcaption\label{figs1:a}
\phantomsubcaption\label{figs1:b}
\phantomsubcaption\label{figs1:c}
\phantomsubcaption\label{figs1:d}
\phantomsubcaption\label{figs1:e}
\phantomsubcaption\label{figs1:f}
\phantomsubcaption\label{figs1:g}
\phantomsubcaption\label{figs1:h}
\phantomsubcaption\label{figs1:i}
\phantomsubcaption\label{figs1:j}
\phantomsubcaption\label{figs1:k}
\phantomsubcaption\label{figs1:l}
\phantomsubcaption\label{figs1:m}
\end{center}
\vskip -0.2in
\end{figure}

\begin{itemize}
    \item Another interesting finding of our simulation was a high Pearson($\Delta$) performance of the mean baseline for truly affected genes under a very low number of perturbations (around $0.6$ in the lower $k$ limit on Supplemental Fig.~\ref{figs1:k}).
    \item We hypothesize this is explained by the sparsity of the true biological differences when perturbations occur in non-overlapping genes. As exemplified on Supplemental Fig.~\ref{figs1:j} for a single perturbation, if this perturbation uniquely shows up-regulation of gene 1 and gene 2 and we are under the low $k$ regime, then the mean baseline $\mu^{all}$ will pick up some signal from it and correlating $r(\Delta^p, \Delta^{all})$ will yield a positive results. This behavior is direct consequence of Pearson($\Delta$) focusing on direction changes rather that dynamic range. Note that because of the low probability of gene perturbation in simulation and sparsity of biological signal in the real data, this behavior is the rule rather than the exception.
    \item Confirming our result and explanation when sub-sampling perturbations in the real data under 10 different seeds we get the same trend when analyzing Pearson ($\Delta$) only on DEGs vs the Rest of perturbations which are a proxy of the real affected genes and are the ones that make every perturbation different from every other (Supplemental Figs. \ref{figs1:l} and \ref{figs1:m}).
    \item As the number of perturbations in the dataset increases the probability of overlap between perturbed genes increases while also the pulling effect of a single perturbation  on the mean baseline is significantly reduced. In other words $\mu^{all}$ is closer to the origin of the plot in Supplemental Fig.~\ref{figs1:j} reducing artificial performance inflation of the mean baseline as observed with simulations and real data.
\end{itemize}

\section{Supplemental figures}
\label{supplemental_figures}

\begin{figure}[ht]
\vskip 0.2in
\begin{center}
\centerline{\includegraphics[width=0.9\textwidth]{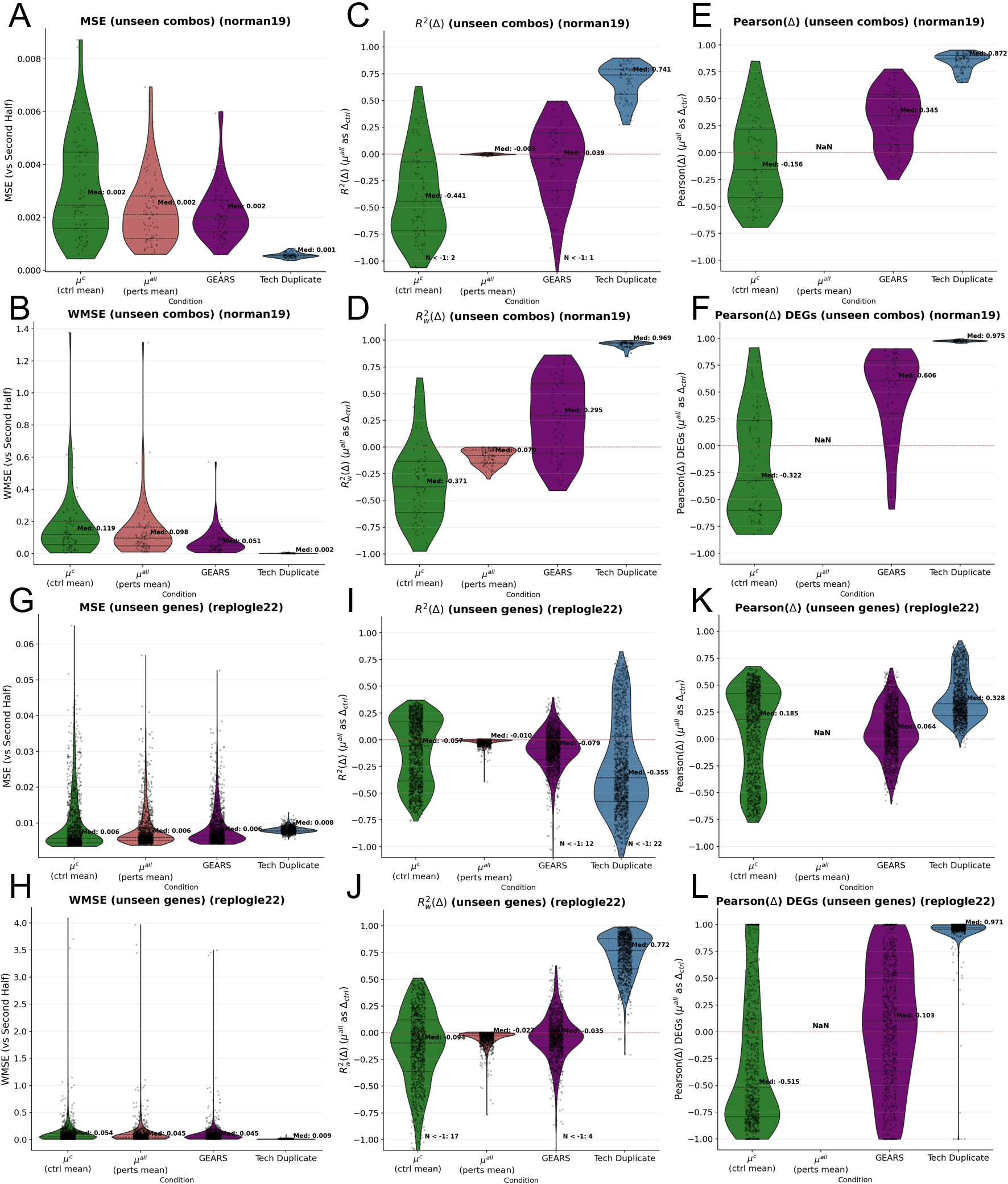}}
\caption{DEG-aware metrics, elimination of control bias, and addition of negative and positive baselines provide greater sensitivity and calibration to assess perturbation response model performance. (A) In the \textit{Norman19} dataset, MSE between ground truth and baselines or model predictions. $\mu^c$ (control cell mean), $\mu^{all}$, GEARS model predictions, and technical duplicate baseline are shown. (B) Same as (A) but measuring error with DEG score-weighted MSE (WMSE) instead of MSE. (C-D) Same as (A-B) but using $R^2(\Delta^p,\hat{\Delta}^p)$ and DEG score-weighted $R^2_w(\Delta^p,\hat{\Delta}^p)$ as the error metric, where $\hat{\Delta}^p$ is $\hat{\mu}^p - \mu^{all}$ and $\Delta^{p}$ is $\mu^{p} - \mu^{all}$. (E-F) Same as (C-D) except with Pearson instead of $R^2$ and filtering for DEGs (per perturbation) instead of weighting by DEG score. (G-L) Same as (A-F) but for the \textit{Replogle22} dataset in which the task was prediction of unseen single genes.
}
\label{figs3}
\phantomsubcaption\label{figs3:a}
\phantomsubcaption\label{figs3:b}
\phantomsubcaption\label{figs3:c}
\phantomsubcaption\label{figs3:d}
\phantomsubcaption\label{figs3:e}
\phantomsubcaption\label{figs3:f}
\phantomsubcaption\label{figs3:g}
\phantomsubcaption\label{figs3:h}
\phantomsubcaption\label{figs3:i}
\phantomsubcaption\label{figs3:j}
\phantomsubcaption\label{figs3:k}
\phantomsubcaption\label{figs3:l}
\end{center}
\vskip -0.2in
\end{figure}

\begin{figure*}[ht]
\vskip 0.2in
\begin{center}
\centerline{\includegraphics[width=\textwidth]{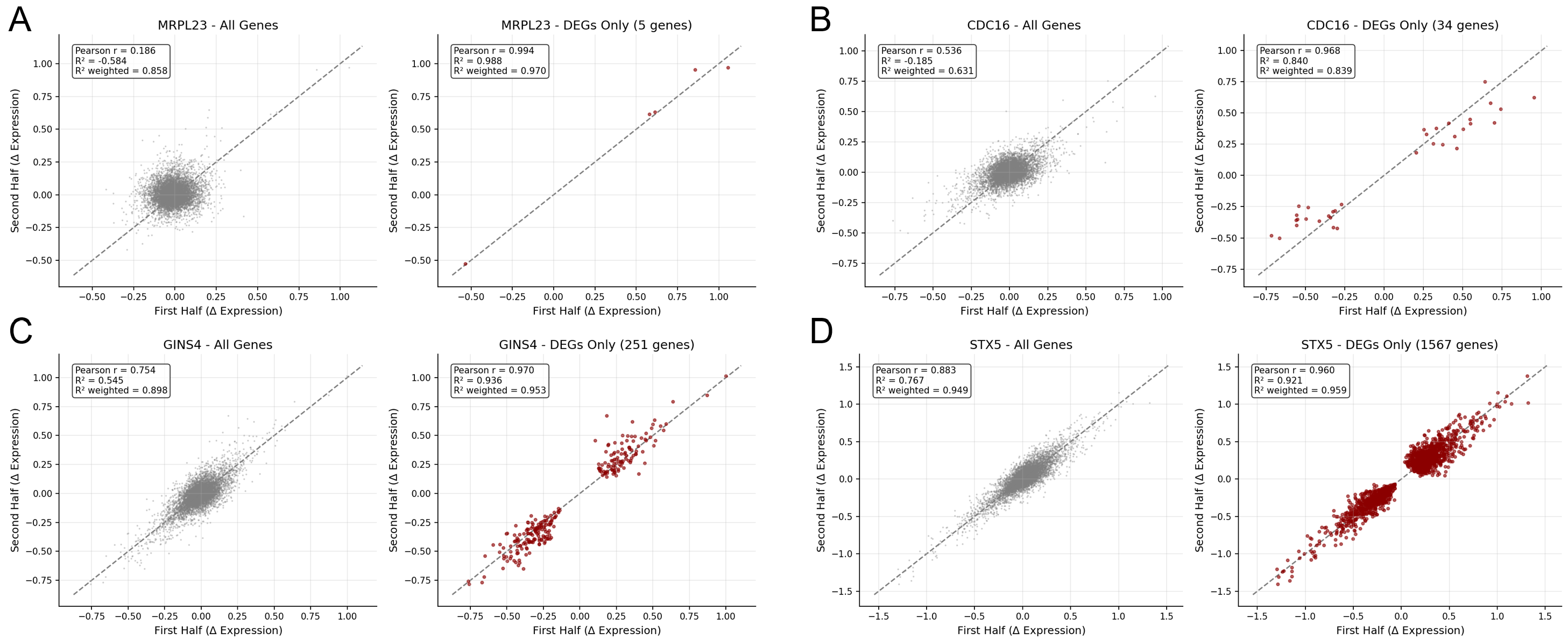}}
\caption{Effect of perturbation strength (measured by number of significant DEGs) on $R^2_w(\Delta^p,\hat{\Delta}^p)$ metric in technical duplicate baseline. (A) Scatter plots showing correlation between perturbation effects ($\Delta = \mu^{p} - \mu^{all}$) when using the first half of the data to predict the second half for MRPL23 with and without filtering for only MRPL23-specific DEGs. DEG-weighted and regular $R^2$ shown, along with Pearson correlation. (B) Same as (A) for CDC16. (C) Same as (A) for GINS4. (D) Same as (A) for STX5.
}
\label{scatterplot-r2}
\phantomsubcaption\label{scatterplot-r2:a}
\phantomsubcaption\label{scatterplot-r2:b}
\phantomsubcaption\label{scatterplot-r2:c}
\phantomsubcaption\label{scatterplot-r2:d}
\end{center}
\vskip -0.2in
\end{figure*}

\begin{figure*}[ht]
\vskip 0.2in
\begin{center}
\centerline{\includegraphics[width=\textwidth]{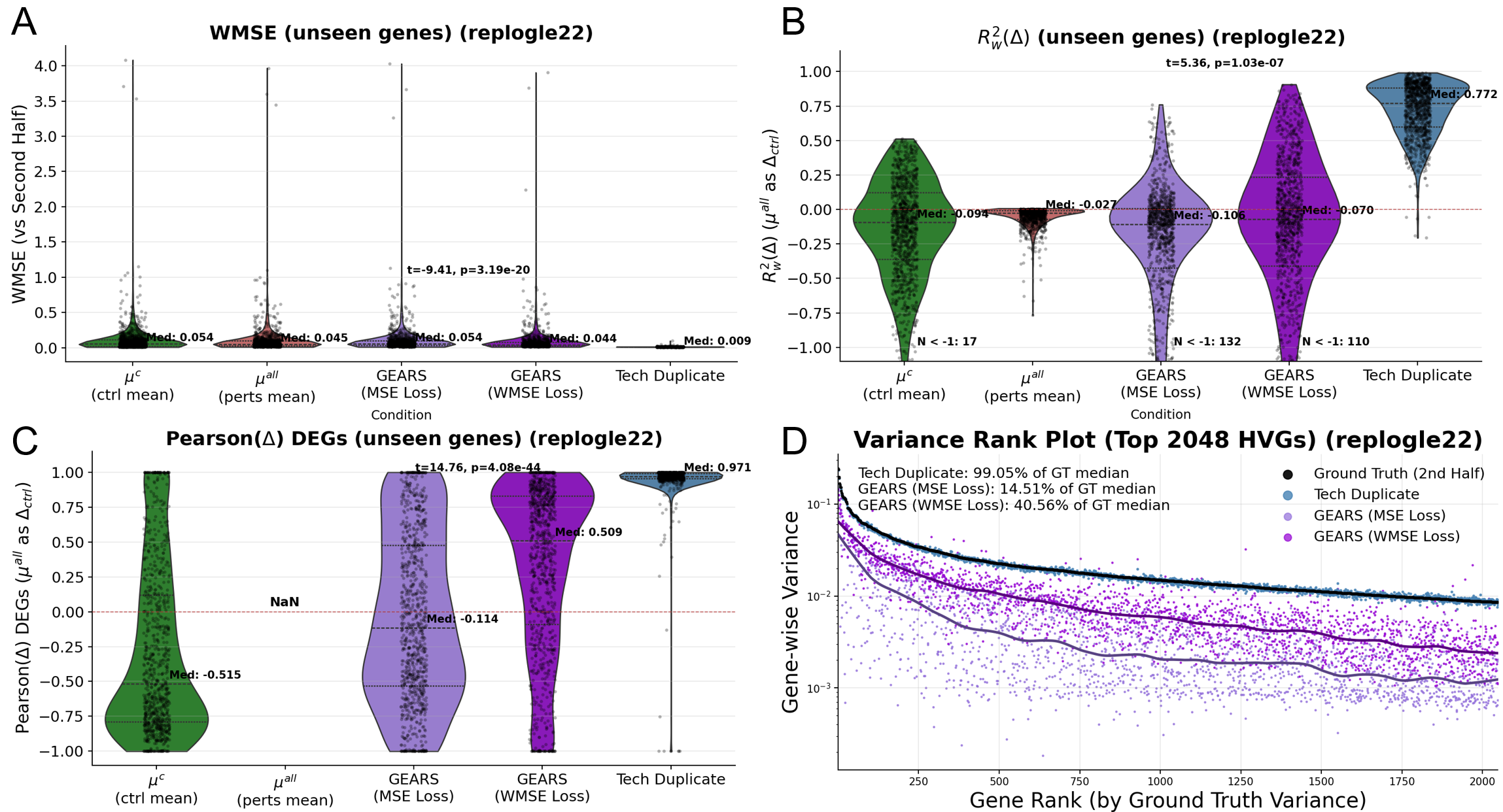}}
\caption{Supplemental figure accompanying Fig.~\ref{wmse-improvement-norman}. DEG score-weighted loss reduces mode collapse and improves model performance (\textit{Replogle22}). (A) WMSE between prediction and ground-truth perturbation mean ($\mu^{p}$). X labels: $\mu^c$ (control mean), $\mu^{all}$ (mean of all perturbed cells), predictions from GEARS model with MSE or WMSE loss, and technical duplicate baseline. Means between GEARS MSE/WMSE compared with paired t-test. (B) Same as (A) but for $R^2_w(\hat{\Delta}^p,\Delta^{p})$, the DEG score-weighted $R^2$ between predicted ($\hat{\Delta}^p$) vs ground-truth perturbation effect ($\Delta^{p}$). For $\Delta$ calculations, $\mu^{all}$ is the reference. (C) Same as (B) but with Pearson correlation and filtering to only include perturbation-specific DEGs (vs Rest). (D) Plot showing the top 2048 highly-variable genes ranked by variance in the ground truth pseudobulked dataset. Includes variances for the Technical Duplicate, GEARS MSE/WMSE predictions.
}
\label{figs4}
\phantomsubcaption\label{figs4:a}
\phantomsubcaption\label{figs4:b}
\phantomsubcaption\label{figs4:c}
\phantomsubcaption\label{figs4:d}
\end{center}
\vskip -0.2in
\end{figure*}

\begin{figure*}[ht]
\vskip 0.2in
\begin{center}
\centerline{\includegraphics[width=\textwidth]{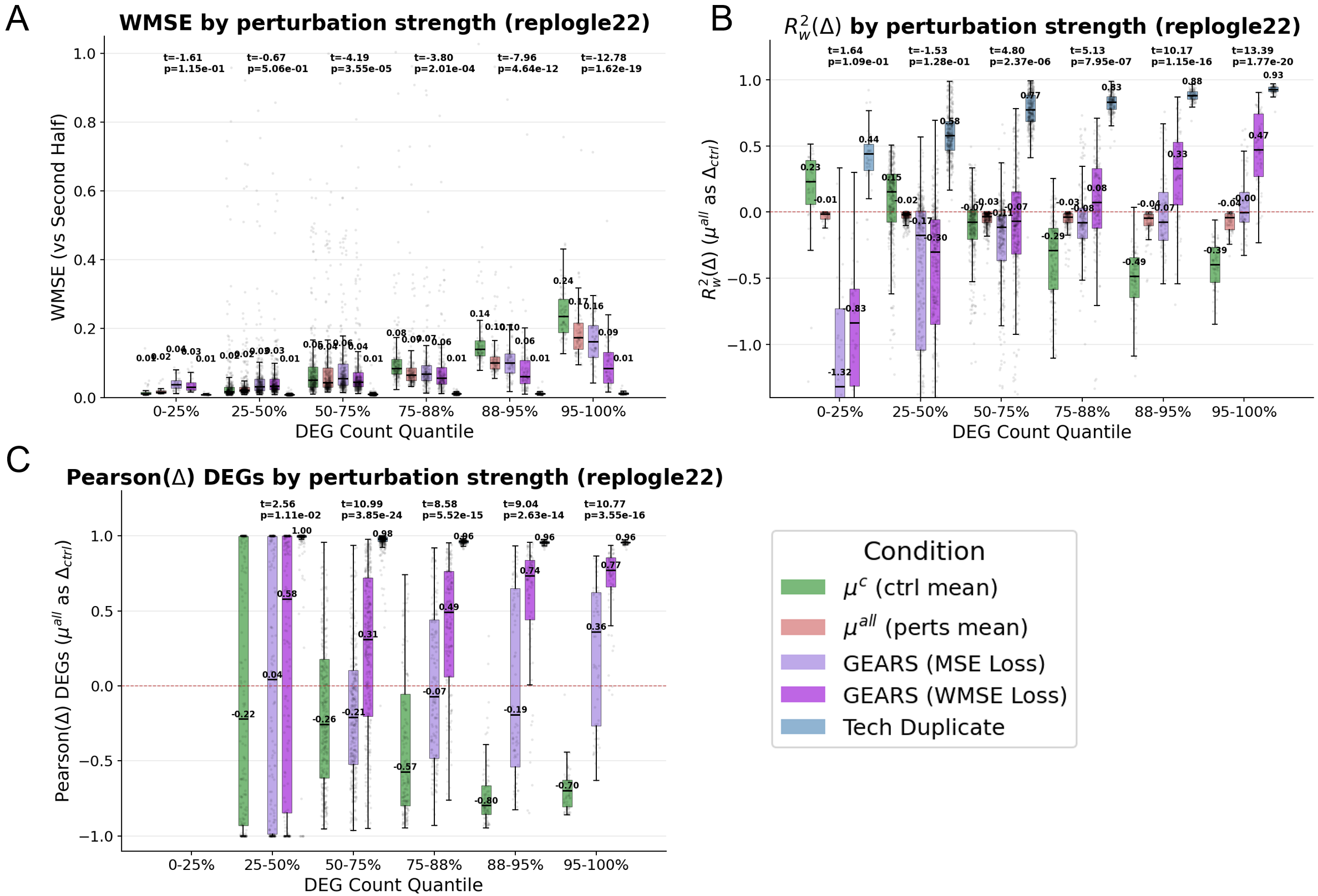}}
\caption{DEG score-weighted loss improves model performance on unseen gene prediction task, especially for stronger perturbations. (A) Performance of baselines and GEARS perturbation prediction (with MSE or WMSE loss), compared using WMSE metric vs ground truth perturbation mean ($\mu^{p}$), grouped by the quantile range of perturbations tested (quantile ranges based on number of DEGs for each perturbation). Paired t-test conducted for each quantile range between GEARS with MSE vs GEARS with WMSE loss, with t-score and p value shown on plot. Median of each prediction within each quantile range also shown. (B) Same as (A) but for $R^2_w(\Delta^p,\hat{\Delta}^p)$ (DEG score-weighted R2 between predicted vs ground-truth perturbation effect). (C) Same as (B) but for Pearson correlation with data filtered to only include perturbation-specific DEGs (vs Rest). Note that the 0-25\% quantile is missing because there were no DEGs for perturbations in this quantile.
}
\label{figs6}
\phantomsubcaption\label{figs6:a}
\phantomsubcaption\label{figs6:b}
\phantomsubcaption\label{figs6:c}
\end{center}
\vskip -0.2in
\end{figure*}


\end{document}